\documentstyle[11pt,aaspp4,psfig]{article}

\begin{document}     

\def\today{\ifcase\month\or
January\or February\or March\or April\or May\or June\or
July\or August\or September\or October\or November\or December\fi
\space\number\day, \number\year}



\newcommand{\squig}{$\sim$}
\newcommand{\squigleq}{\mbox{$^{<}\mskip-10.5mu_\sim$}}
\newcommand{\squiggeq}{\mbox{$^{>}\mskip-10.5mu_\sim$}}
\newcommand{\squiggeqmm}{\mbox{$^{>}\mskip-10.5mu_\sim$}}
\newcommand{\decsec}[2]{$#1\mbox{$''\mskip-7.6mu.\,$}#2$}
\newcommand{\decsecmm}[2]{#1\mbox{$''\mskip-7.6mu.\,$}#2}
\newcommand{\decdeg}[2]{$#1\mbox{$^\circ\mskip-6.6mu.\,$}#2$}
\newcommand{\decdegmm}[2]{#1\mbox{$^\circ\mskip-6.6mu.\,$}#2}
\newcommand{\decsectim}[2]{$#1\mbox{$^{\rm s}\mskip-6.3mu.\,$}#2$}
\newcommand{\decmin}[2]{$#1\mbox{$'\mskip-5.6mu.$}#2$}
\newcommand{\asecbyasec}[2]{#1$''\times$#2$''$}
\newcommand{\aminbyamin}[2]{#1$'\times$#2$'$}
\newcommand{\Lra}{$\Longrightarrow$}

\title{Empirical Uncertainty Estimators for Astrometry from Digital Databases}
\author{Eric W. Deutsch}
\affil{Department of Astronomy, 
       University of Washington, Box 351580,
       Seattle, WA 98195-1580\\
       deutsch@astro.washington.edu}

\begin{center}
Accepted for publication in the Astronomical Journal\\
To appear in the 1999 October issue\\
{\it received 1999 April 7; accepted 1999 June 8}
\end{center}


\begin{abstract}

In order to understand the positional uncertainties of arbitrary
objects in several of the current major databases containing
astrometric information, a sample of extragalactic radio sources with
precise positions in the International Celestial Reference Frame (ICRF)
is compared with the available positions of their optical
counterparts.  The discrepancies between the radio and various optical
positions are used to derive empirical uncertainty estimators for the
USNO-A2.0, USNO-A1.0, Guide Star Selection System (GSSS) images, and
the first and second Digitized Sky Surveys (DSS-I and DSS-II).  In
addition, an estimate of the uncertainty when the USNO-A2.0 catalog is
transferred to different image data is provided.  These optical
astrometric frame uncertainties can in some cases be the dominant error
term when cross-identifying sources at different wavelengths.

\end{abstract}


\clearpage
\section{INTRODUCTION}

In the past decade it has become increasingly easier to tie an
arbitrary optical field to a standard astrometric reference frame with
arcsecond precision.  With the recent publication of the USNO-A2.0
astrometric catalog (Monet et al. 1998) it is now possible to achieve
\squig\decsec{0}{5} accuracy at any point on the sky.  This greatly
enhances the ability to search for optical counterparts of sources
detected at other wavelengths.  This new accuracy is particularly
desirable when searching for radio sources, which typically have positions
more accurate than is achievable for the optical frame (see, e.g., Deutsch
et al. 1999).  In addition, the {\it Chandra X-ray Observatory} will
for the first time provide X-ray positions with arcsecond accuracy,
creating a need for better optical precision.  Indeed, even if the
internal precision of a radio or X-ray position is well understood, it
is often difficult to obtain a quantitative estimate of the additional
uncertainty introduced by the transfer of these coordinates to a given
optical astrometric frame.  For many problems, this component can be the
dominant source of positional error.  This work describes an empirical,
uniform test of the astrometric accuracy of several well known databases
containing astrometric information.

\subsection{{\it HST} GSC and GSSS Images}

The first astrometric catalog with full and dense sky coverage was the
{\it Hubble Space Telescope} ({\it HST}) Guide Star Catalog (GSC),
which is described in detail by Lasker et al. (1990).  Some initial
defects in version 1.0 of this catalog have been corrected in the
version 1.1 release.  The GSC contains some $2\times10^7$ objects
incompletely down to $m\sim15$.  As it is possible that a deep, small
field CCD image may contain few, if any, unsaturated GSC stars, it is
often necessary to refer directly to the Guide Star Selection System
(GSSS) images, the plate data used to create the GSC itself.  In the
North, 1982--1985 epoch ``Quick~V'' plates are used, and in the South,
deeper 1974--1984 epoch UK SERC J plates are used.  Each plate image is
digitized to \decsec{1}{7} pixel$^{-1}$, and the coefficients for a third
order polynomial astrometric solution are included in the image header,
defining the GSC astrometric frame.  The astrometric accuracy of the
GSSS images is discussed by Russell et al. (1990).  Although the GSC
has been available on CD-ROM for a decade, extractions from the GSSS
images themselves have only been accessible outside the ST\,ScI via
the WWW (http://archive.stsci.edu/) for the last few years.  Note that
additional astrometric improvements are implemented in the GSC 1.2
release (see R\"{o}ser et al. 1997) but as these improvements are not
easily incorporated into the GSSS images, this catalog is not considered
here.

\subsection{DSS-I}

The next astrometric archive to become widely available was the Digitized
Sky Survey (DSS-I).  This image dataset was distributed on 102 CD-ROMs
after undergoing a lossy $10\times$ compression process.  In the South,
the UK SERC J GSSS images are used, while in the North, a digitization
of the  1950--1956 epoch Palomar Observatory Sky Survey I (POSS-I) E
(red) plates is used, again at \decsec{1}{7} pixel$^{-1}$.  Although the
astrometric accuracy in the South is identical to that of the {\it HST}
GSC, the astrometric accuracy of the DSS-I in the North is acknowledged
to be inferior to the GSC.  The POSS-I E data were selected for the
DSS-I because they are deeper than the ``Quick V'' data, and comparably deep
to the southern UK SERC J images.  The DSS-I was a major milestone as it
made a snapshot of the sky down to $m\sim20$ with embedded
astrometric information available to anyone with access to the CD-ROM
set, or shortly thereafter, anyone with WWW access to the ST\,ScI or
other sites where the Survey data are available on-line.  An assessment
of the astrometric accuracy of the DSS-I has been previously published by
V\'eron-Cetty \& V\'eron (1996), and a more detailed description of the
DSS-I itself can also be found therein.

\subsection{DSS-II}

The Palomar Observatory Sky Survey II (POSS-II) and the Second Epoch
Survey (SES) in the south are currently being processed to form the
Second Digitized Sky Survey (DSS-II).  It contains a 1985-1997 epoch
multicolor entire-sky survey using newer, better quality plates digitized
to slightly more than \decsec{1}{0} pixel$^{-1}$.  While nearly all of
the data have been obtained, only about 85\% of the sky (red plates only)
is available at the ST\,ScI WWW interface at the time of this writing,
and it should be further emphasized that these images are being made
available with prepublication status.  The astrometric information
assessed here is likely to change when the final product is available.
Recent status reports of the Digitized POSS-II (DPOSS; does not include
the SES) have been presented by Djorgovski et al. (1997, 1998).

\subsection{USNO-A1.0}

The USNO-A1.0 astrometric catalog (Monet et al. 1996) contains
$5\times10^8$ objects down to $m\sim20$, detected on plates
digitized by the Naval Observatory's Precision Measuring Machine
(PMM).  For $\delta>-33^\circ$, POSS-I O and E plates are used, and
for $\delta<-33^\circ$, UK SERC-J and European Southern Observatory
ESO-R survey plates are used.  Only objects which are detected in
{\it both} the red and blue plates (within 1$''$) are included in the
catalog, and the catalog positions are derived from the blue plates.
Both a blue and a red magnitude are also published for each object in
this catalog, which is available in a 10 CD-ROM set and via the WWW.
The absolute astrometric frame is based on the {\it HST} GSC version
1.1, and is therefore not significantly more accurate than the GSC.
Whereas the images discussed above come only with a single polynomial
astrometric solution to describe the entire plate, the USNO catalog
positions are based on images which are described with a polynomial
plus a correction map to compensate for the systematic errors inherent
at the edges of the Schmidt plates.  This allows significantly better
precision over the entire area of the plates.  See also Canzian (1997)
for additional information.

\subsection{USNO-A2.0}

The USNO-A2.0 astrometric catalog (Monet et al. 1998) is based on the
USNO-A1.0 with a few important differences.  The most important
difference is that the absolute astrometric calibration is based on the
International Celestial Reference Frame (ICRF) as realized by the USNO
ACT catalog (Urban et al. 1998), which itself is based on the Tycho
catalog (ESA 1997), generated from the Hipparcos mission.  Another
important difference is the source of individual positions; whereas the
A1.0 gives blue plate positions, the A2.0 gives the average of the blue
and red plate positions.  POSS-I O and E plates were typically obtained
on the same night, but SERC-J and ESO-R plates were obtained at
different epochs and thus the catalog position will place objects with
proper motion at the average epoch of the two plates; objects which
have moved more than 1$''$ between the two epochs will not be included
in the catalog at all.   The alignment with the ICRF makes the
USNO-A2.0 the most accurate astrometric catalog of faint stars, i.e.
the stars likely to be found unsaturated on an arbitrary CCD image.
This catalog is available on an 11 CD-ROM set and via the WWW from the
US Naval Observatory Flagstaff Station (http://www.nofs.navy.mil/).
The image data used, which was digitized at about \decsec{0}{9}
pixel$^{-1}$, may also become remotely available in early 2000.

\subsection{GSC-II}

Although not yet available at this time, the Second Guide Star Catalog
(GSC-II) will likely be available soon, and promises to provide positions
comparably accurate to the USNO-A2.0, as well as magnitudes, colors
and proper motions.  This catalog will be based on the DSS-II plate
material, and the first epoch for the proper motion measurements will
be based on the DSS-I plates.  The GSC-II is merely mentioned here as a
future valuable resource for completeness and is not discussed further.
For additional information on the construction and status of the GSC-II,
see McLean et al. (1997, 1998).

\section{ANALYSIS}

To assess the astrometric accuracy of these databases, a dataset of
objects with extremely accurate radio positions is obtained from Ma et
al. (1998), who present a careful derivation of VLBI positions, most of
which have milliarcsecond (mas) accuracy, for 608 extragalactic radio
sources in the ICRF.  These sources are grouped into three classes,
of which all are used in this work.  The handful of sources which have
positional uncertainties larger than 30 mas are not used.  The sources
are fairly evenly distributed over the entire sky.  The right ascension
zero point has been aligned to the ICRF which is consistent with the
FK5 J2000.0 optical frame.

For each of these sources, a small \aminbyamin{1}{1} region about each
position is extracted from the USNO-A2.0 catalog and the DSS-I images.
The images are examined individually, and those optical counterparts which
are not suitable for this study are thrown out.  Briefly, 62 sources are
somewhat subjectively deemed too faint for an accurate centroid based on
the DSS-I image (not quite the same data as used to create the catalog,
however); 64 sources are too crowded by neighbors or too extended;
136 sources were too faint and not even in the USNO-A2.0 catalog; 63
sources were not present in the USNO-A2.0 catalog but yet appeared to
have sufficient signal in the DSS-I images that they probably ought to
have been.  This leaves a dataset of 283 sources for which the data are
acceptable in at least some of the databases discussed here.  The datasets
used for testing of the various databases are sometimes reduced further.
For the three imaging databases, a minimum of 20,000 ADU (``counts'')
within a $5''$ aperture is required; for example this reduces the dataset
used for the GSSS images by 41 sources as the northern ``Quick V''
images are not as deep as the DSS-I images.  Two of the 283 sources are
not in the USNO-A1.0 catalog (in fact those particular fields contain
far fewer sources in the A1.0 than A2.0 overall).  As noted before,
not all of the DSS-II images are yet available on-line, and thus the
accuracy of the DSS-II was only assessed with the 235 sources for which
a good DSS-II detection was available.

For the 283 sources, positions from all the available databases, either
directly from a catalog or centroided on an image, are recorded along
with the radio coordinates of Ma et al. (1998) for later inspection.
The total flux of the sources is also measured using simple aperture
photometry within a $5''$ radius.

In the fields of $\sim100$ of the radio sources, a set of isolated,
easily measurable reference stars with USNO $R<18$ is selected to
transfer the USNO-A2.0 astrometric frame to each of the three images.
The transfer fit usually employs $30-80$ stars, depending on the field
density, and yields an uncertainty in the transfer of $<$\decsec{0}{05}.
The transferred solution is a simple linear solution (sufficient for
the small \aminbyamin{10}{10} fields used for this purpose) derived
with a least-squares fit using procedures from the IDL {\it Astronomy
User's Library} (Landsman 1993) and written in IDL by the author.
The coordinates of each optical counterpart are then also recorded
using the transferred solution.

\section{DISCUSSION}

Using the optical positions of these extragalactic radio sources measured
in these various ways, the distribution of the deviation from the radio
positions is examined.  In Table 1 relevant derived parameters from these
distributions are listed for the different databases.  Each astrometric
database is listed in Column 1.  Columns 2--3 list the average
$\Delta\alpha$ and $\Delta\delta$, indicating any global shift from
the ICRF.  In columns 4 through 6 the maximum observed $\Delta\alpha$,
$\Delta\delta$, and $\Delta$R ($\Delta{\rm R}=\sqrt{\Delta\alpha^2 +
\Delta\delta^2}$) errors in arcseconds within the best 68 percent of
the respective distributions are given.  The next four sets of 3 columns
list similar values within the 90, 95, 99, and 100 percentiles, where
the latter means all of the sources.  The number of sources examined
for each database is provided in the last column.  For example, for
the USNO-A2.0, 90\% of the 283 objects examined differed radially
from the radio position by less than or equal to \decsec{0}{40}.
Similarly, 90\% of the 283 objects (but not necessarily the same 90\%
as above) differed in $\delta$ by less than or equal to \decsec{0}{30}.
These results will be discussed for the various databases in turn.

\subsection{USNO-A2.0}

The USNO-A2.0 is clearly superior to the other frames considered here.
The average $\Delta\alpha$ and $\Delta\delta$ are quite small but
yet deviate from 0.0 in a marginally significant way.  The astrometric
uncertainties are better than the other databases by nearly a factor of 3,
with $1\sigma\approx\decsecmm{0}{25}$, 90\% confidence at \decsec{0}{40},
and even $3\sigma\approx\decsecmm{0}{5}$.  In fact, the largest observed
discrepancy in this sample of 283 sources is \decsec{0}{63}!  This is
quite remarkable given the possibility of true misalignment between
a radio source and a slightly extended optical galaxy, not readily
discernible as extended on the low spatial resolution DSS-I images.
A brief examination of this most discrepant source, 0458--020, yields
no anomalies in the images and is cataloged as a high-$z$ QSO, and
thus not likely to suffer from intrinsic radio/optical misalignment.
Centroiding errors probably also contribute to these overall uncertainties
in addition to local frame offset errors, but it is the combined effect
that it most relevant in most cases.

In Fig. 1 the distribution of discrepancies is plotted against a
variety of other parameters.  The first panel depicts the discrepancies
in $\Delta\alpha$ and $\Delta\delta$ in arcseconds, in the sense
that the radio coordinate is subtracted from the optical coordinate.
The distribution is quite tight and slightly offset from the center.
A fiducial \decsec{0}{5} radius circle is overplotted.  The second panel
shows the combined (radial) discrepancy ($\Delta$R) vs. Declination; no
Declination region is apparently worse than others, even near the poles.
The third panel shows $\Delta$R vs. the distance of the source from the
plate center on the DSS-I images, which is mostly the same plate material
as used to create the USNO-A2.0.  The units are in \decsec{1}{7} pixels
as encoded in the image headers.  The horizontal solid line is simply
the \decsec{0}{5} fiducial.  The vertical dashed line indicates the
approximate useful half width of the plates.  Clearly the astrometric
solution is essentially as good near the edges of the plate as near
the center.

The fourth panel shows $\Delta$R vs. the USNO $R$ magnitude.  It does
appear that the faintest stars post slightly worse values.  If all stars
$R>18$ are excluded, the resulting errors are $\sim10\%$ better for the
new sample of 225 objects (see Table 1).  In the fifth panel $\Delta$R
is plotted vs. the USNO $(B-R)$ color.  Even the bluest and reddest
objects show no obviously greater errors than the others.  Finally in
the last panel $\Delta$R is plotted vs. $\log_{10}ADU$ within a $5''$
aperture around the source in the DSS-I image; this is not directly
indicative of the brightness in the images used to create the A2.0,
but a reasonably good estimate of it.  A similar figure is generated
for each of the databases examined.  They are not all reproduced here,
but are available from the author.

Figure 2 displays the distributions of $\Delta\alpha$ and $\Delta\delta$
in arcseconds for the USNO-A2.0 283 object sample.  The slight average
offsets evident in Table 1 can also be readily seen here and appear to
be significant.  The reason for this systematic offset remains unclear.
Certainly for program objects of color significantly different from the
typical USNO-A2.0 star, atmospheric differential refraction must play
some role.

\subsection{USNO-A1.0}

The USNO-A1.0 catalog is worse than the A2.0 by nearly a factor of 3,
and shows similar results as the GSSS images, not surprising as the
A1.0 is based on the absolute astrometric calibration of the GSSS/GSC.
The results are probably slightly better due to improved plate
distortion corrections.  A set of plots like Fig. 1 for this database
turns up nothing unusual.

\subsection{GSSS Images}

The GSSS images provide the next best astrometric database.  It is
still in use for {\it HST} observation planning.  Russell et al. (1990)
assessed the astrometric accuracy of the GSC and GSSS images.  Using a
small sample of 48 extragalactic radio sources, they derived
$\sigma_\alpha=\decsecmm{0}{63}$, $\sigma_\delta=\decsecmm{0}{58}$,
although the values were better in the north.  However, with their
much larger ``CAMC'' sample, they find for the whole plate
$\sigma_\alpha=\decsecmm{0}{58}$, $\sigma_\delta=\decsecmm{0}{53}$,
and slightly better,
$\sigma_\alpha=\decsecmm{0}{55}$, $\sigma_\delta=\decsecmm{0}{48}$,
for the inner 50\% of plates.  This compares
favorably to the
$\sigma_\alpha=\decsecmm{0}{47}$, $\sigma_\delta=\decsecmm{0}{47}$,
$\sigma_{\rm R}=\decsecmm{0}{71}$ values derived here, and lends
considerable confidence that the reduction described herein is
accurate.

\subsection{DSS-I}

The DSS-I begins with a reasonably good $1\sigma$ astrometric uncertainty
but degrades at $>99$\% confidence to the worst statistics of the group.
The worst offset (for the object 0202+319) is \decsec{4}{6}.  The accuracy
in the South is identical to that of the GSSS images, as they are the
same data.  In the North, the POSS E images have inferior astrometric
fit quality, as is noted in the documents which come with the CD-ROMs,
although it is perhaps not widely appreciated how far off the astrometry
can be.  A figure like Fig. 1 for this dataset indicates that the worst
discrepancies lie near the outer edges of the plates.  If all sources
greater than 4500 pixels (\decdeg{2}{125}) from the plate centers are
excluded, the sample of 144 remaining objects shows a somewhat better
distribution as seen in Table 1.  Clearly, positions derived from the
DSS-I should be treated with caution as there is a potential to be several
arcseconds in error in the North.  A far more detailed discussion of
the DSS-I astrometric accuracy is presented by V\'eron-Cetty \& V\'eron
(1996).  They derive similar uncertainties as in Table 1 and derive
some corrections.

\subsection{DSS-II}

Immediately noticeable in the DSS-II row in Table 1 are the large
$<$$\Delta\alpha$$>$, $<$$\Delta\delta$$>$.  In the plot like Fig. 1a
for the DSS-II it is immediately evident that most sources exhibit a
($+1'',-1''$) offset.  As the pixelsize is $\approx1''$ it is likely
that this is due to a one pixel offset somewhere in the pipeline which
generates the astrometric solution for these images.  It is, however,
also readily evident that a smaller subset of discrepancies do lie about
(0,0) and thus some images may not suffer from this defect.  For this
reason, it may not be prudent to simply apply this offset to the images.
A second set of statistics is generated for the DSS-II with all source
($\alpha, \delta$) positions adjusted by ($-1'', +1''$).  The result is
better, but still not as good as the GSSS image or USNO catalogs.  It
should be stressed that these results are derived from the
prepublication data available at this time, and the astrometry in the
final product may be significantly improved.

\subsection{Transferring the USNO-A2.0 Frame to Images}

The uncertainties involved in obtaining an ICRF position for an
extragalactic point source from the USNO-A2.0 catalog is now fairly
well established.  The positions of objects in the Galaxy with known
proper motion can also readily determined, although determining the exact
epoch of the A2.0 positions is nontrivial (the epoch for each field is
available in a separate catalog at the NOFS WWW site).  For objects with
unknown proper motion, the uncertainties must be estimated on a case by
case basis, depending on the data available.

The remaining case is for objects simply not listed in the USNO-A2.0
catalog, for whatever reason.  Since large telescopes with modern
detectors can go much fainter than the A2.0 is a relatively short time,
this is actually a frequent issue.  Obtaining coordinates for such
objects is easily done by fitting an astrometric solution to the new
image using an input list of USNO-A2.0 ($\alpha, \delta$) and image (X, Y)
positions for a set of stars common to both the catalog and the image.
Given a few tens of reference stars, the final uncertainty in the
transferred solution due to centroiding errors can easily be less than
\decsec{0}{05}.  However, if the epoch of the USNO-A2.0 and image data
differ significantly, the small proper motions of the reference stars will
dominate the uncertainty.  Stars with large proper motions can, of course,
be easily rejected from the fit.  If the directions of motions of the
reference stars are mostly random, the residuals are merely inflated.
The worst scenario is where the tiny proper motions of many reference
stars yield a small systematic shift due to Galactic rotation.  This can
be beneficial for stars moving with the reference
stars, but is detrimental to determining positions of extragalactic
objects, or objects otherwise not associated with the bulk motion.
In an effort to quantify the uncertainties in determining positions of
extragalactic objects not available in the USNO-A2.0, an A2.0 solution is
transferred to several of the database images used here and the resulting
position of the radio source counterpart measured as described in \S 2.
The results for the DSS-II images may be most relevant as they have
fairly recent epochs and are better sampled and deeper than the other
images used here.

The results of this effort are tabulated in the last three rows of
Table 1, where column 1 lists the source database which is aligned to
the USNO-A2.0.  The results yield a $\sim25$\% degradation over the
A2.0 itself, which is quite respectable given that the difference
in epoch between A2.0 images and DSS-II images can be up to 40 years.
For many applications these larger values will be a better estimate of
the true uncertainty of measured positions based on, but not directly
from, the USNO-A2.0.

In closing the reader is reminded that many precise positions are still
found in B1950, and special care must be taken when converting them to
J2000 if subarcsecond accuracy is important.  Some generic precession
utilities do not perform the necessary FK4 to FK5 correction, and the
resulting improperly converted positions may be in error by $\sim1''$.

\acknowledgments

Thank you to Bruce Margon, Scott Anderson, and Brian McLean for their many
helpful suggestions for this work, and to David Monet and collaborators
for making the USNO catalogs available to the community.  Thanks also
to Claire Vayssiere for pointing out the DSS I astrometry discrepancy
in an earlier version of this manuscript.  Support for this work was
provided by NASA through grant number AR-07990.01-96A from the ST\,ScI,
which is operated by AURA, Inc.

\bigskip

The Digitized Sky Surveys were produced at the Space Telescope Science
Institute under U.S. Government grant NAG W-2166. The images of these
surveys are based on photographic data obtained using the Oschin Schmidt
Telescope on Palomar Mountain and the UK Schmidt Telescope. The plates
were processed into the present compressed digital form with the
permission of these institutions.

The National Geographic Society - Palomar Observatory Sky Atlas (POSS-I)
was made by the California Institute of Technology with grants from the
National Geographic Society.

The Second Palomar Observatory Sky Survey (POSS-II) was made by the
California Institute of Technology with funds from the National Science
Foundation, the National Geographic Society, the Sloan Foundation,
the Samuel Oschin Foundation, and the Eastman Kodak Corporation.

The Oschin Schmidt Telescope is operated by the California Institute
of Technology and Palomar Observatory.  The UK Schmidt Telescope was
operated by the Royal Observatory Edinburgh, with funding from the UK
Science and Engineering Research Council (later the UK Particle Physics
and Astronomy Research Council), until 1988 June, and thereafter by the
Anglo-Australian Observatory. The blue plates of the southern Sky Atlas
and its Equatorial Extension (together known as the SERC-J), as well
as the Equatorial Red (ER), and the Second Epoch [red] Survey (SES)
were all taken with the UK Schmidt.

All data are subject to the copyright given in the copyright
summary. Copyright information specific to individual plates is provided
in the downloaded FITS headers.  Supplemental funding for sky-survey
work at the ST\,ScI is provided by the European Southern Observatory.


\clearpage
\oddsidemargin -0.8in

{\tiny
\begin{deluxetable}{lrr|rrr|rrr|rrr|rrr|rrr|r}
\tablenum{1}
\tablewidth{8.0in}
\tablecolumns{19}
\tablecaption{Empirical Uncertainty Estimates in Arcseconds Relative to the ICRF}
\tablehead{
\colhead{} &
\colhead{} &
\colhead{} &
\colhead{} &
\colhead{68\%} &
\colhead{} &
\colhead{} &
\colhead{90\%} &
\colhead{} &
\colhead{} &
\colhead{95\%} &
\colhead{} &
\colhead{} &
\colhead{99\%} &
\colhead{} &
\colhead{} &
\colhead{100\%} &
\colhead{} &
\colhead{} \\
\cline{4-6}
\cline{10-12}
\cline{16-18}
\colhead{Database} &
\colhead{$<$$\Delta\alpha$$>$} &
\colhead{$<$$\Delta\delta$$>$} &
\colhead{$\Delta\alpha$} &
\colhead{$\Delta\delta$} &
\colhead{$\Delta$R} &
\colhead{$\Delta\alpha$} &
\colhead{$\Delta\delta$} &
\colhead{$\Delta$R} &
\colhead{$\Delta\alpha$} &
\colhead{$\Delta\delta$} &
\colhead{$\Delta$R} &
\colhead{$\Delta\alpha$} &
\colhead{$\Delta\delta$} &
\colhead{$\Delta$R} &
\colhead{$\Delta\alpha$} &
\colhead{$\Delta\delta$} &
\colhead{$\Delta$R} &
\colhead{N} 
}
\startdata 
USNO-A2.0         & $ 0.04$ & $ 0.04$ &  0.17 & 0.18 & 0.26 &  0.29 & 0.30 & 0.40 &  0.34 & 0.34 & 0.45 &  0.45 & 0.43 & 0.52 &  0.57 & 0.59 & 0.63 & 283 \nl
USNO-A2.0 ($R<18$)& $ 0.03$ & $ 0.03$ &  0.16 & 0.17 & 0.24 &  0.27 & 0.26 & 0.36 &  0.32 & 0.34 & 0.42 &  0.43 & 0.43 & 0.52 &  0.45 & 0.45 & 0.53 & 225 \nl
USNO-A1.0         & $-0.11$ & $-0.12$ &  0.47 & 0.41 & 0.65 &  0.78 & 0.67 & 1.04 &  0.94 & 0.85 & 1.24 &  1.43 & 1.23 & 1.46 &  1.67 & 1.41 & 1.70 & 281 \nl
GSSS              & $-0.12$ & $-0.07$ &  0.47 & 0.47 & 0.71 &  0.84 & 0.78 & 1.10 &  1.08 & 0.99 & 1.32 &  1.43 & 1.31 & 1.66 &  1.68 & 1.37 & 1.72 & 242 \nl
DSS-I             & $-0.08$ & $ 0.03$ &  0.53 & 0.48 & 0.75 &  1.02 & 0.88 & 1.26 &  1.22 & 1.14 & 1.56 &  1.63 & 2.37 & 2.68 &  2.13 & 4.59 & 4.59 & 274 \nl
DSS-I ($r<\decdegmm{2}{125}$)    & $-0.10$ & $-0.04$ &  0.46 & 0.37 & 0.65 &  0.87 & 0.66 & 1.07 &  0.81 & 1.08 & 1.40 &  1.63 & 1.37 & 1.68 & 1.68 & 2.12 & 2.68 & 144 \nl
DSS-II             & $ 0.74$ & $-0.93$ &  1.07 & 1.16 & 1.60 &  1.43 & 1.61 & 1.95 &  1.60 & 1.82 & 2.16 &  2.04 & 2.41 & 2.54 &  2.28 & 3.07 & 3.08 & 235 \nl
DSS-II ($-1,+1$)   & $-0.26$ & $ 0.07$ &  0.63 & 0.52 & 0.85 &  1.10 & 0.97 & 1.43 &  1.23 & 1.21 & 1.66 &  1.80 & 1.57 & 2.35 &  2.17 & 2.07 & 2.42 & 235 \nl
A2.0 \Lra GSSS    & $ 0.02$ & $ 0.09$ &  0.20 & 0.24 & 0.33 &  0.31 & 0.38 & 0.48 &  0.39 & 0.45 & 0.56 &  0.66 & 0.54 & 0.69 &  0.68 & 0.55 & 0.71 & 101 \nl
A2.0 \Lra DSS-I     & $ 0.02$ & $ 0.03$ &  0.18 & 0.19 & 0.32 &  0.34 & 0.36 & 0.44 &  0.38 & 0.38 & 0.51 &  0.47 & 0.53 & 0.56 &  0.68 & 0.54 & 0.69 & 104 \nl
A2.0 \Lra DSS-II   & $ 0.04$ & $ 0.13$ &  0.21 & 0.26 & 0.35 &  0.36 & 0.36 & 0.45 &  0.43 & 0.41 & 0.53 &  0.48 & 0.53 & 0.59 &  0.52 & 0.53 & 0.67 & 108 \nl
\enddata
\end{deluxetable}
}

\clearpage
\oddsidemargin 0in

\begin{figure}
\plotone{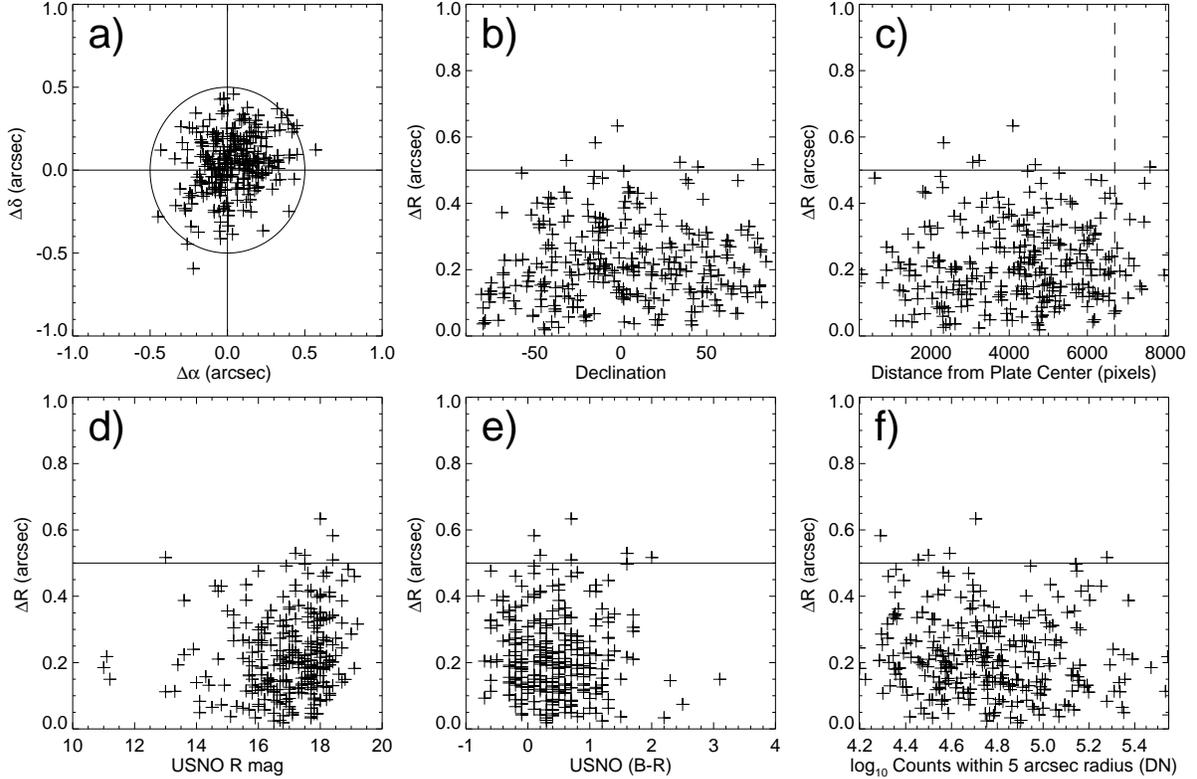}
\caption{The distribution of discrepancies between USNO-A2.0 and VLBI
radio positions (Ma et al. 1998) for the sample of 283 extragalactic
radio sources.  A fiducial \decsec{0}{5} radius circle or horizontal
line is overplotted in each panel.
{\it a}) $\Delta\alpha$ vs. $\Delta\delta$ in arcseconds;
{\it b}) $\Delta$R (radial discrepancy) in arcseconds vs. Declination;
{\it c}) $\Delta$R in arcseconds vs. distance from the plate center
  (\decsec{1}{7} pixels) where the dashed vertical line denotes the
  plate half width;
{\it d}) $\Delta$R in arcseconds vs. USNO-A2.0 $R$ magnitude;
{\it e}) $\Delta$R in arcseconds vs. USNO-A2.0 $(B-R)$ color;
{\it f}) $\Delta$R in arcseconds vs. $\log_{10}ADU$ within a $5''$ radius
   aperture on the DSS-I images.
There do not appear to be any large systematic trends in any of these
representations.}
\end{figure}

\begin{figure}
\plotone{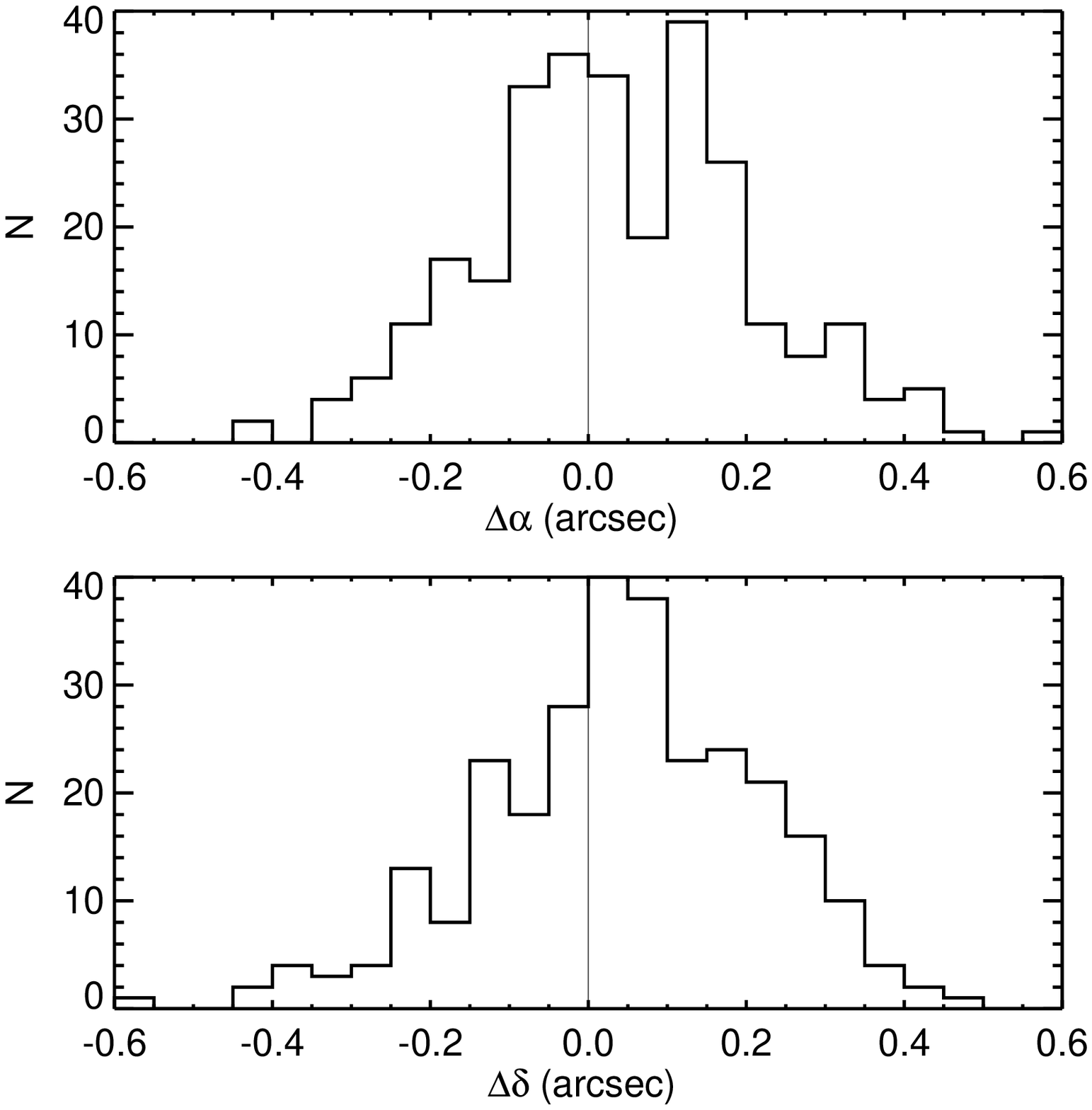}
\caption{Histograms of $\Delta\alpha$ and $\Delta\delta$ discrepancies
between USNO-A2.0 and radio positions for a sample of 283 extragalactic
radio sources.  There do appear to be slight zeropoint offsets in the two
distributions, although their origin and significance remains unclear.
Certainly color terms must play some role.}
\end{figure}


\clearpage
\begin{thebibliography}{}

\bibitem[]{} Canzian, B. 1997, in New Horizons from Multi-Wavelength Sky
Surveys, ed. B. J. McLean, D. A. Golombek, J. J. E. Hayes, \& H. E. Payne
(Dordrecht: Kluwer), 422

\bibitem[]{} Deutsch, E. W., Margon, B., Anderson, S. F., Wachter, S. \&
Goss, M. 1999, ApJ, 524, in press

\bibitem[]{} Djorgovski, S. G., De Carvalho, R. R., Gal, R. R.,
Pahre, M. A., Scaramella, R., \& Longo, G. 1997, in New Horizons
from Multi-Wavelength Sky Surveys, ed. B. J. McLean, D. A. Golombek,
J. J. E. Hayes, \& H. E. Payne (Dordrecht: Kluwer), 424

\bibitem[]{} Djorgovski, S. G., Gal, R. R., Odewahn, S. C., Brunner,
R. J., \& de Carvalho, R. R. 1998, BAAS, 30, 1270

\bibitem[]{} ESA 1997, The Hipparcos and Tycho Catalogues, SP-1200

\bibitem[Landsman 1993]{lan93} Landsman, W. B. 1993, in ASP Conf. Ser.
52, Astronomical Data Analysis Software and Systems II, ed. R. J.
Hanisch, R. J. V.  Bissenden, \& J. Barnes (San Francisco: ASP), 256

\bibitem[Lasker et al.,\ 1990]{las90} Lasker, B. M., Sturch, C. R.,
McLean, B. J., Russell, J. L., Jenkner, H., \& Shara, M. M. 1990, \aj,
99, 2019

\bibitem[]{} Ma, C., Arias, E. F., Eubanks, T. M., Fey, A. L., Gontier,
A.-M., Jacobs, C. S., Sovers, O. J., Archinal, B. A., Charlot, P. 1998,
AJ, 116, 516

\bibitem[]{} McLean, B., Hawkins, G., Spagna, A., Lattanzi, M., Lasker,
B., Jenkner, H., \& White, R. 1997, in New Horizons from Multi-Wavelength
Sky Surveys, ed. B. J. McLean, D. A. Golombek, J. J. E. Hayes, \&
H. E. Payne (Dordrecht: Kluwer), 431

\bibitem[]{} McLean, B., Lasker, B., Lattanzi, M. 1998, BAAS, 30, 900

\bibitem[]{} Monet et al. 1996, USNO-A1.0: A Catalog of Astrometric
Standards, U.S. Naval Observatory

\bibitem[]{} Monet et al. 1998, USNO-A2.0: A Catalog of Astrometric
Standards, U.S. Naval Observatory

\bibitem[]{} R\"{o}ser, S., Morrison, J., Bucciarelli, B., Lasker, B.,
\& McLean, B. 1997, in New Horizons from Multi-Wavelength Sky Surveys,
ed. B. J. McLean, D. A. Golombek, J. J. E. Hayes, \& H. E. Payne
(Dordrecht: Kluwer), 420

\bibitem[Russell et al.,\ 1990]{rus90} Russell, J. L., Lasker, B. M.,
McLean, B. J., Sturch, C. R., \& Jenkner H. 1990, \aj, 99, 2059

\bibitem[]{} Urban, S. E., Corbin, T. E., \& Wycoff, G. L. 1998, The ACT
Reference Catalog, U.S. Naval Observatory

\bibitem[]{} V\'eron-Cetty, M.-P., \& V\'eron, P. 1996, A\&AS, 115, 97

\end{thebibliography}
\end{document}